\documentclass[12pt,preprint,epsf]{aastex}

\def\spose#1{\hbox to 0pt{#1\hss}} 
\def\simlt{\mathrel{\spose{\lower 3pt\hbox{$\mathchar"218$}}      
\raise 2.0pt\hbox{$\mathchar"13C$}}} 
\def\simgt{\mathrel{\spose{\lower 3pt\hbox{$\mathchar"218$}}      
\raise 2.0pt\hbox{$\mathchar"13E$}}} 
  
\def\etal{{\rm et~al.~}} 
\usepackage{epstopdf}

\lefthead{Hislop \et al} 
\righthead{Extragalactic Distance Scale } 
\begin{document}
\title{The Extragalactic Distance Scale without Cepheids IV.}

\author{Lachlan Hislop \& Jeremy Mould} 
\affil{School  of Physics, University of Melbourne, Vic 3010, Australia} 
\authoraddr{E-mail: jmould@unimelb.edu.au} 
\author{Brian Schmidt, Michael S. Bessell, Gary Da Costa, Paul Francis, Stefan Keller, Patrick Tisserand, Sharon Rapoport, \& Andy Casey}
\affil{Research School of Astronomy \& Astrophysics, Australian National University}
\authoraddr{E-mail: brian@mso.anu.edu.au} 


\begin{abstract}
The Cepheid period-luminosity relation is the primary distance indicator used in most determinations of the Hubble constant. The tip of the red giant branch (TRGB) is an alternative basis. 
Using the new ANU SkyMapper Telescope, we calibrate the Tully Fisher relation
in the I band. We find that
the TRGB and Cepheid distance scales are consistent.
\end{abstract}

\section{Introduction}
The extragalactic distance scale of the Hubble Space Telescope (HST)
Key Project \citep{ke95} is based on the Cepheid period-luminosity (PL) relation
and secondary distance indicators, such as the Tully-Fisher relation \citep{
sa00}, the supernova
standard candle \citep{gib00}, surface brightness fluctuations \citep{fe00}, and the fundamental plane
\citep{ke00}. It has been criticized recently \citep{{str08},{str07},
{st08}}
on the grounds that the PL relation may not be unique. Indeed, the finite width
of the Cepheid instability strip in the HR diagram implies that nuisance parameters
such as metallicity, helium abundance, and star formation history may play a role in determining the PL
relation. Metallicity was considered as a second parameter by \cite{fr01},
\cite{sa04}, and \cite{mac06}, and linear corrections were made based
on measured values of [O/H] from the galaxies' HII regions. 

The classical extragalactic distance scale continues to be important because measurements
of the Hubble Constant with WMAP \citep{la10} are model dependent: they integrate
the scale factor all the way to high redshift. 

It is of interest, therefore, to see how well the distance scale can be measured
without reference to Cepheids at all. In Papers I, II, and III \citep{{ms08},{ms09},{ms09b}} we
calibrated the H-band Tully-Fisher relation, surface brightness fluctuations, the Fundamental Plane, and Type Ia supernovae using the tip of the red giant branch (TRGB) and found a distance scale compatible with that of the H$_0$ Key Project 
\citep{mo00}. In this paper we use the TRGB
distance indicator to examine the I-band Tully-Fisher relation. As discussed in Paper I, the TRGB is a good
standard candle because it results from the helium flash on the red giant branch, which
theory suggests is relatively immune to metallicity effects in old stellar populations. \cite{ri07} find hundredth of a magnitude systematics in TRGB
for old metal poor stellar populations. \cite{sg05} show that
this dispersion grows substantially for young and metal rich populations.

\section{The Calibration Sample}

The calibration sample from which a Tully-Fisher relation will be constructed consists of 13 galaxies, which are presented in Table 1. This sample is limited to galaxies that have consistent methods by which their distances, I-band apparent magnitudes, and rotational velocities are measured. Each of these criteria is outlined below. 

\subsection{Distance Moduli}

TRGB distance moduli for the calibration galaxies were taken from the Extragalactic Distance Database (EDD)\footnote{http://edd.ifa.hawaii.edu/} and are listed in Table 1. The procedure for obtaining distance moduli is described in \cite{Jacobs09}. CMDs for the galaxies were produced from HST observations, which then allows the apparent magnitude of the TRGB to be measured following detection using the maximum likelihood algorithm of \cite{MAK06}. TRGB distance moduli can then be obtained using the absolute magnitude of the TRGB as calibrated by \cite{ri07}, and are expressed as:
\begin{equation}
\mu = m_{\mathrm{TRGB}} - A_{I} +4.05 -0.217[(V-I) - (A_{V}-A_{I}) - 1.6]
\end{equation}
for the Johnson-Cousins I and V filters, which have been corrected for galactic extinctions A$_{I}$ and A$_{V}$. No corrections have been made for extinction within the host galaxy, but the observed RGB populations have been chosen such that they lie in the outer halo, where dust is scarce. For these distance moduli, 1$\sigma$ total errors are presented.\\

There have been multiple measurements of the TRGB distance modulus for many of the individual calibration galaxies, but we adopt the EDD values due to their consistent calibration.

\subsection{Consistency of Distance Moduli}
Calibration of the Tully-Fisher relation using TRGB distance moduli cannot be performed without first assessing the consistency of distance measurements to nearby galaxies from different methods. Firstly, we perform a two-fold consistency check between the EDD listed values for the TRGB distance moduli and Cepheid-based distances, as well as additionally determined TRGB measurements. We do this by plotting in Figure 1 histograms of the independently determined Cepheid and TRGB distance moduli for each galaxy, along with the EDD measurement. It can be seen that consistency to within 0.2 mag is achieved for most calibrator galaxies.\\

It can also be seen for NGC 3368, NGC 3621 and NGC 3627, that these $\mu_{\mathrm{TRGB}}$ measurements represent distinct lower estimates for their respective distances. Furthermore, the adopted EDD distance modulus for NGC 3351 differs from the smallest published Cepheid measurement by a further 0.5 mag, and differs even more significantly from the $\mu_{\mathrm{TRGB}}$ published in \cite{ri07}, which marks it as a clear outlier. We found that the origin of this large difference between the distance moduli of \cite{Jacobs09} and \cite{ri07} is the difference between I-band apparent magnitude measurement of the TRGB itself, published values of which differ by $\sim$ 1.3 mag. While this may indicate a possible discrepancy in the method by which the observed magnitude of the TRGB is measured, in both instances the apparent magnitude of the TRGB has been measured following maximum likelihood method of \cite{MAK06}. We do not expect perfect agreement between TRGB and Cepheid distances where, as per \cite{sa04}, differences between these two measurements provides an estimate for the metallicity dependence of the Cepheid PL relation. Furthermore, we still expect some variation about the derived Cepheid distances, as each will be slightly offset from different measurements that have been calibrated differently (e.g. using Cepheid populations in either the LMC or the maser galaxy NGC 4258). \\

In light of this uncertainty, we nevertheless choose to adopt $\mu_{\mathrm{TRGB}} = 29.92$ as published in \cite{ri07}, since this exhibits a more realistic discrepancy between Cepheid and TRGB-based distance measurements. \\

A similar discrepancy between TRGB distance moduli measurements exists for the Antennae galaxies (NGC 4038/39), where the values published by \cite{schw08} and \cite{sav08} differ by 0.9 mag. Schweizer \etal suggest that the difference is likely due to the contaminating presence of AGB stars, which led Saviane \etal to  misidentify the TRGB in the CMD. It is possible that this issue of contamination may also apply to measurements of the TRGB for NGC 3351.\\

When considering consistency between different measurements of the TRGB distance modulus, $\mu_{\mathrm{TRGB}}$, it is important to note that methods by which the tip itself is detected differ between groups, with recently employed methods being the maximum likelihood approaches of \cite{MEZ02} and \cite{MAK06}, and the Gaussian-smoothed Sobel edge-detection filter used by \cite{BEL01} and \cite{sa04}. While it is beyond the scope of this paper to determine the reliability of one method over another, it is imperative that we adopt distance measurements that have been consistently calibrated. We have favored more recent calibrations of M$^{TRGB}_{I}$ in light of criticisms from \cite{ri07} that a significantly more populated CMD is required to accurately detect the TRGB than those used by \cite{LFM93}. And while \cite{MF95} suggest a minimum of $\sim$ 100 stars for accurate detection, this number has been revised to the far more conservative 400 - 500 stars suggested by \cite{Madmag09}. \citeauthor{Madmag09} have attempted to resolve these statistical limitations by expressing the I-band apparent magnitude of the TRGB as a function of color (and hence metallicity), such that individual stars can be corrected for color and essentially transferred to the same TRGB magnitude. The calibration of M$^{TRGB}_{I}$ by \citeauthor{Madmag09} agrees well with that of \citeauthor{ri07}.  \\

Interestingly, \cite{Dalcan09} have published multiple TRGB distance moduli for a number of different fields for a range of nearby galaxies as part of the ACS Nearby Galaxy Survey Treasury (ANGST). Detections of the TRGB are determined using a combination of techniques from \cite{MEZ02} and \cite{SMF96}.  Where the measurement of M$^{TRGB}_{I}$ differs by less than 0.02 mag for each field within a galaxy, we are able to briefly assess the extent to which the choice of field within a galaxy to measure the TRGB magnitude affects the yielded distance modulus. We expect neither variations in metallicity nor internal extinction amongst halo RGB populations, where metallicities here are typically low and internal extinction is assumed to be negligible. The extent of crowding and hence contamination of the CMB by AGB stars may differ slightly between fields, but again we do not expect the populations of the RGB fields to differ significantly enough to render measurements of the TRGB inconsistent between fields. Upon inspection of the results for a number of galaxies, distance moduli for each galaxy vary at most by 0.1 mag, which are consistent to within errors for an uncertainty of 0.05 mag, which suggests that any variations can be largely attributed to random photometric errors. Uncertainties introduced by adopting the same foreground reddening value for each field may also contribute to the dispersion of distance moduli measurements \citep{ri07}, although this suggestion has not been fully supported.

\subsection{Photometry}
The total sample of photometric data has been sourced from SDSS data and our own observations performed using the ANU SkyMapper Telescope, along with data taken from the literature.

\subsubsection{SkyMapper Photometry} 
Photometry was obtained for four galaxies from the calibration sample (NGC 55, NGC 247, NGC 253, and NGC 7793) from observations performed using the ANU SkyMapper 1.35m telescope \citep{K07}, located at Siding Spring Observatory, on September 23 and 24, 2010. Observations were performed in the $g$ and $i$ bands and the images were bias-corrected and flat-fielded using bias frames and twilight flats taken on the night. These were applied using the IRAF $imarith$ package. Integration times were set at 60 s for all objects. \\

For each galaxy, calibration from an $i$-band magnitude to a standard I-band was performed using observations of standard stars from \cite{Grah82}.  

\subsubsection{SDSS Photometry}
Photometry for galaxies NGC 3351 and NGC4826 was obtained from the SDSS catalogue\footnote{http://das.sdss.org/www/html/}, and was separately calibrated using the included observations of Landolt standards \citep{Land92} . \\

The calibration of SkyMapper and SDSS data is discussed in Appendix 1. Photometry for the remaining calibration galaxies (NGC 300, NGC 891, NGC 2403, NGC 3368, NGC 3621, NGC 3627, and NGC 4258) was obtained from \cite{TP00}.\\


\subsubsection{Integrated Magnitude Profiles and Comparison of IRAF and ARCHANGEL Outputs}
Apparent magnitudes were measured using both the IRAF \textit{ellipse} routine, and the ARCHANGEL photometry package developed by James Schombert\footnote{http://abyss.uoregon.edu/$\sim$js/archangel/}. ARCHANGEL performs the core routines of sky determination, ellipse fitting, and surface brightness profile generation which are common to all photometry software packages. Following \citet{Schomb07}, the procedure for measuring apparent magnitudes is outlined below. \\

Images are first cleaned of foreground contaminating objects, and a preliminary maximal radius out to which ellipses are fit is then determined. This allows for accurate sky background determination that avoids both foreground objects and the outer regions of the galaxy. In order to measure the background, boxes of a set size are placed randomly in the image, and a mean count is obtained for each box. The final magnitude of the background is then taken as the mean of each of these mean sky box values.\\ 

Isophotal contours are then fit to the image as ellipses, which simply act as tracers for the stellar material. The quality of these can be manually assessed by observing rapid changes in parameters such as position angle and eccentricity towards the outermost ellipses. Once ellipses are fitted, a surface brightness profile is produced and a disk fit can be applied to yield apparent magnitudes.\\

Magnitudes are evaluated at $-2.5$ times the logarithm of the sum of the pixel values within each fitted ellipse from which the total sky background inside the ellipse is subtracted. A final raw magnitude for the galaxy is arrived at where the integrated magnitude values within each successive ellipse converge, signifying that additional pixel values encompassed by the largest fitted ellipse are purely sky background. Thus, where these plots converge (see Figures 2 and 3) signifies the luminous boundary of the galaxy.\\

The availability of the IRAF and ARCHANGEL fitting routines allowed us to verify the output of each program, particularly with regard to the integrated magnitudes. Difficulties arose in this particular instance of using images of nearby galaxies, where a consistent set of elliptical apertures could not be fit to the highly resolved (and in the case of NGC 55, highly asymmetrical) galactic structure. Images were either re-scaled or smoothed until a successful fit was achieved in either IRAF or ARCHANGEL. IRAF ellipse fits were successfully applied to all 6 calibrators for which photometry was available, and ARCHANGEL fits were successful on 3 of the 6. Comparisons of the fits for galaxies with both IRAF and ARCHANGEL fits are plotted in Figure 3.\\

Agreement between the IRAF and ARCHANGEL integrated magnitudes was achieved for NGC 7793 to within 0.05 mag, to within 0.15 mag for NGC 4826, and to within 0.4 mag for NGC 55. Given that measurement errors on apparent magnitudes are typically $\sim$ 0.05 mag, a disagreement of 0.1 mag and above is certainly less than ideal. This disagreement in magnitude values can of course be attributed to the differences in the fitting algorithms that are employed by IRAF and ARCHANGEL. Indeed, even by visual inspection it is clear that ellipses are being fit out to different radii in each output. Although we have no real reason to doubt the validity of either fitting routine, one way to verify the output from both IRAF and Archangel is to visually inspect the ellipse fits and compare them to a visual estimate of the boundary of the galaxy. This is certainly not ideal since the entire point of ellipse fitting routines is to avoid such visual estimates. Nevertheless, we expect difficulty with these images given that most objects have their very outer edges clipped by the physical size of the CCD chip, along with the diffuse nature of a galaxy's outer disk, which is compounded by the high sky brightness under which the observations were performed.\\


The difficulties encountered are particularly evident for NGC 247, whose integrated magnitude curve exhibits no convergence for the fitted ellipses. We can attribute this result to the diffuse nature of this dwarf galaxy, where the fitting algorithms struggled to define an outer luminous boundary. For this particular object, we assign a larger error of 0.1 mag.\\

For all remaining galaxies with a single fit in either IRAF or ARCHANGEL, we visually inspect the ellipse fits to ensure they have appropriately encompassed the galaxy, and assign a larger error estimate where there is a clear discrepancy between the fitted ellipses and a visual estimate of the fit.

\subsection{Photometric Corrections}
The output from photometry programs such as ARCHANGEL represents an initial, raw measurement for galactic apparent magnitudes. 
This is corrected by $k$ times the airmass where
$k$ is the atmospheric extinction coefficient, which must be determined for each telescope site \citep{HungBess00}. 
Measurement errors on the observed magnitudes are typically given by the propagated uncertainty in the sky background measurement, which becomes the dominant source of error where the faint edges of the galaxy are difficult to determine \citep{Schomb07}. Sky backgrounds are determined using the \textit{skybox} routine in ARCHANGEL, which places boxes throughout a frame, albeit away from the galaxy and foreground stars, and determines a mean sky value for each box. The final sky value for the image is then taken as the mean of each box mean, and the uncertainty taken as the 1/$\sqrt{n}$ of the rms dispersion. We then apply a calibration offset to each of these measured magnitudes, as outlined in Appendix 1, and propagate the uncertainty in the calibration offset along with the aforementioned uncertainty to determine a final measurement error.\\

Measured apparent magnitudes are then corrected for foreground Galactic extinction using the E(B$-$V) redenning values of \cite{Schlegel98}, which serves as an estimate for the amount of light from distant objects that is scattered by dust within our own galaxy along its line of sight. Magnitudes were k-corrected using k$_{I} = 0.16z$, as per \cite{Han92}.\\

Magnitudes are also corrected for internal extinction using $\Delta$m$_{i}$ = $-\gamma$log($b/a$) as per \cite{Giov94}, where $b/a$ is the axial ratio of the galaxy - the ratio of the minor and major axes. This internal extinction correction provides an estimate for an inclination dependent extinction, whereby light emitted within galaxies that are inclined edge-on ($i \sim 90^{\circ}$) undergoes a larger amount of internal scattering relative to galaxies that are inclined face-on ($i \sim 0^{\circ}$). Methods of internal extinction corrections are numerous, and do not agree as to how the $\gamma$ parameter should be parametrised (see \citealt{sa00} for outline). Furthermore, some methods will incorporate a morphology-dependent offset, whereas others will not. The correction used by \cite{Giov94} is perhaps ideal then, since it ignores the additional morphology dependence, in light of the criticism from \citeauthor{sa00} regarding the uncertainties involved in morphology classification. We have chosen to evaluate the $\gamma$ parameter as per \cite{Tully98} in favor of \citeauthor{Giov94}.\\

The final expression for the corrected apparent magnitudes is that given by \cite{G97a}:
\begin{equation}
m_{c} = m_{obs} - A_{I} + k_{I} - \Delta \mathrm{m}_{i}
\end{equation}

Following \citeauthor{G97a}, the total measurement error on the corrected apparent magnitudes is the quadrature sum of the measurement error and the error associated with the internal extinction correction. The error on the internal extinction correction is given by the propagation of the errors in $\gamma$ and axial ratio $b/a$.


\subsection{Line Widths}

For our calibration sample galaxies, we adopt the rotational velocity measurements presented in \cite{Court09}, a database of homogeneously determined line widths using the profiles taken from HI catalogues published by \cite{Korib04}, \cite{Spring05}, \cite{Hucht05}, \cite{Theureau06}, \cite{Giov07}, \cite{Saint08}, and \cite{Kent08}. \citeauthor{Court09} define a window over the profile which excludes a small portion of the integrated flux in the wings, such that a total flux can be determined within that window, and a mean flux taken over the number of spectral channels. The line widths are then estimated at 50\% of the mean flux density per channel.\\

For galaxies with multiple profiles, a line width is measured for each available profile. For consistency, we adopt line widths derived from profiles taken from the same source as the cluster sample (\citeauthor{Spring05}), but this is not possible for our calibration sample. For most of the calibration sample galaxies, \citeauthor{Court09} line widths have been measured using the either the HIPASS BGC profiles (\citealt{Korib04}), or from profiles presented in \cite{Spring05}. For NGC247, a W$_{m50}$ line width was determined using the profile presented in \cite{Carpu90}. For NGC 891 and NGC 2403, profiles were taken from \cite{Rots1980}.\\

Measurement uncertainties for each line width are limited to the spectral resolution after smoothing. \citeauthor{Court09} provide conservative error estimates, and apply a threshold error of 20 km s$^{-1}$ to indicate galaxies suitable for distance applications, in line with \cite{TP00}.\\

Other possible line width measurements are listed in \cite{Spring05} and \cite{G97a}, such as widths measured as 50\% of the peak flux, $W_{50}$. Where consistency between the calibration and cluster samples is preserved, the choice of line width is largely irrelevant and should be motivated by breadth and accuracy of available data. We choose not to use the line widths included in the SFI++ sample, as we are unable to apply the same correction for instrumental effects to our calibration sample line widths.  

\subsubsection{Line Width Corrections}
In accordance with \cite{sa00}, line widths are corrected for inclination and redshift.
\begin{equation}
W_{c} = \frac{W_{obs}}{(\mathrm{sin}i)(1+z)}
\end{equation} 

We forgo the corrections for instrumental effects and turbulence since this cannot be applied consistently across both data sets. Errors in the corrected line widths are given by the propagation of the measurement error on the observed line width and the error in the inclination. 

\subsection{Inclinations}
One of the key parameters in Tully-Fisher applications is the axial ratio, $b/a$, from which an inclination is derived, which allows apparent magnitudes to be corrected for internal extinction, and to project line-of-sight rotational velocities. Following our ellipse fitting routines, the axial ratio is determined at the I $\sim$ 20 mag arc sec$^{-2}$ level, where our plots of $b/a$ versus ellipse radius converged. Of the 6 galaxies for which we measured inclinations, these values agree with values published in \citet{tp92}, \citet{Spring07} and the RC3 catalogue \citep{devauc95} to within 1-2$^{\circ}$, except for NGC 55 and NGC 3351, where published values vary between 7$^{\circ}$ and 5$^{\circ}$ repsectively. For this study, we have adopted our measured values.

\section{Fitting Procedures and the Tully-Fisher Relation}

For our calibration sample, we apply direct, inverse and bivariate least squares fits as outlined in \cite{G97b}. Each of these approaches assumes a typical straight line fit for the Tully-Fisher relation of the form, $M = a +b \mathrm{log} W$, where $a$ and $b$ are the zero-point and slope respectively, but each fitting method weights the residuals of the merit function\\
\begin{equation}
\chi^{2} = \sum_{i=1}^{N} \left[\frac{M_{i} - M(\mathrm{log}W_{i};a,b)}{\sigma_{i}} \right]^{2}
\end{equation}\\
differently. For the direct fit, residuals in absolute magnitude are minimised such that $\sigma_{i}$ represents the total error in the absolute magnitude. \\

For the inverse fit, line widths are expressed as a function of absolute magnitude, such that the dependent and independent variables have switched roles between the direct and inverse fits. Our model Tully-Fisher relation is now expressed as log $W = M/b - a/b$, and residuals are minimised with respect to the errors in the line width. \\

For the bivariate fit, values for the slope and zero-point are determined according to $M = a +b \mathrm{log} W$ as before, but instead the merit function is now minimised according to orthogonal residuals, where $\sigma_{i}^{2} = \sigma_{M,i}^{2} + b^{2}\sigma_{x,i}^{2}$, such that errors in absolute magnitude and line width are weighted equally. Bivariate fits are performed following the method of \cite{Press92}.\\

\subsection{Commentary on the Fitting Routines}
The use of the direct, inverse, and bivariate fits are not uncommon to Tully-Fisher applications (e.g \citealt{G97b}, \citealt{sa00}), and it is certainly true that all three fitting procedures satisfy the basic function of providing a prediction of a variable when its dependent variable is known. But each regression fit will result in a different measurement of the slope and zero-point in cases where errors are present in both variables and are unique for each data point, which becomes important where H$_{0}$ exhibits particular sensitivity to either of these parameters. Furthermore, \cite{Isobe90} argue that each different fitting method represents a result with theoretically distinct implications, and \cite{FBab92} reiterate that values of H$_{0}$ derived from different fitting techniques are not directly comparable. These concerns are nevertheless somewhat muted in applications of the Tully-Fisher relation, since its origin is purely empirical (see \citealt{TF77} and sources therein) such that insights into the physical relation between absolute magnitude and rotational velocity remain secondary to its practicality.\\

While we are purely interested in using the Tully-Fisher relation to provide absolute magnitudes for galaxies with measured rotational velocities, in which instance we would be quite satisfied in applying the direct fit as a simple predictive tool, we must nevertheless take into consideration the errors in line width when fitting data, as these remain the dominant contribution to the overall error at smaller distances. At larger distances, the statistical uncertainties in the distance dominate. \\

The direct and inverse fits are applied purely for comparison's sake to previous calibrations of the Tully-Fisher relation (e.g, \citealt{sa00},\citealt{TP00}). \cite{Isobe90} contend that these fitting methods are applicable where the cause of the scatter about the fitted relation is unknown, but also note that a direct fit applicable only where the independent variable (rotational velocity in this instance) is measured without error. We adopt the bivariate fitting method as the most robust estimate of the Tully-Fisher parameters, where measurement errors in both absolute magnitude and line width are treated as equal contributers to the overall scatter. \\



\subsection{A TRGB-Calibrated Tully-Fisher Relation}
Figure 4 shows the Tully-Fisher relation derived from 13 calibration galaxies, where the direct, inverse and bivariate fits are superimposed on the single plot, represented by the dot-dashed, dashed, and solid lines respectively. \\

The Tully-Fisher relation for our calibration sample using the bivariate fit is thus:\\
\begin{equation}
M_{I} = (-7.91 \pm 0.62)(\mathrm{log}W - 2.5) - (21.11 \pm 0.07)
\end{equation}\\

Values for the slopes, zero-points and observed scatter for each fit are listed in Table 2. As can be seen, there is a clear discrepancy between the values of the zero-point for the direct and inverse fit when compared with the value for the bivariate fit. But when comparing to the Tully-Fisher parameters of \cite{G97b} derived using the same fitting techniques for similarly sized samples, we can see that such disagreements are not uncommon.\\

\subsection{Scatter about the Tully-Fisher Relation}
We measure the scatter about the Tully-Fisher relation by determining an RMS dispersion of the magnitude residuals, which corresponds to a total observed scatter, $\sigma_{\mathrm{obs}}$. This scatter will have a contribution from the error measurements in apparent magnitude and distance, but what is typically found is that the measurement scatter does not account for the total observed scatter (e.g., \cite{sa00}; \cite{MEY08}). In this way, an intrinsic scatter is implied. The intrinsic scatter is of particular interest as it illustrates an underlying variation between luminosity and rotational motion which can be used to explore how these properties are physically related.\\

Following \cite{MEY08}, an upper limit to the intrinsic scatter can be determined by including in the sample to which a Tully-Fisher relation is fit only galaxies whose measurement errors are effectively minimised. From this reduced sample, the intrinsic scatter can be measured using, $\sigma_{obs}^{2} = \sigma_{meas}^{2} + \sigma_{int}^{2}$. However, the lack of a robust calibration sample for this study means that in addition to our Tully-Fisher relation being poorly constrained, we have only a poor estimate of the intrinsic scatter. Since we require a measure of the intrinsic scatter in order to propagate this uncertainty into a final error on our Tully-Fisher distance moduli, we instead adopt the intrinsic scatter of 0.25 mag via \cite{sa00}, which was taken from their Cepheid-calibrated I-band Tully-Fisher relation. 


\section{Application of the Tully-Fisher Relation to Distant Cluster Samples}

With our fitted Tully-Fisher relation, we now have the I-band absolute magnitude expressed as a function of the logarithm of rotational velocity, such that distances to objects can be obtained where their rotational velocities and apparent brightnesses remain observable.\\

In order to measure H$_0$, we require a large sample of galaxies across a range of distances such that a Hubble diagram, in which an object's distance is plotted against its recessional velocity, can be constructed. However, we are instead interested in measuring distances to local and more distant clusters, rather than individual galaxies. For clusters at a large enough distance, we can safely neglect the physical extent of the cluster and assume that all galaxies assigned to a particular cluster lie at the same distance, where the foreground and background residuals are averaged out. Furthermore, for a cluster containing $n$ suitable galaxies, we have $n$ independent measures of the clusters distance, which allows us to reduce the statistical uncertainty in the cluster's distance by a factor of 1/$\sqrt{n}$.\\

By similarly determining an average recessional velocity for each cluster, we can obtain a value of H$_{0}$ for each cluster. The use of multiple clusters allows us to compensate for cluster motions relative to the Hubble flow, i.e peculiar motion, such that for a distribution of clusters across the sky, these motions can be averaged out. The use of multiple clusters additionally allows the consistency of the value of H$_{0}$ to be assessed over a large range in distance.

\subsection{Outline of the Cluster Data Sample - SFI++ Catalogue}
The SFI++ catalogue \citep{Spring07}, contains I-band photometry along with optical and 21cm line widths of $\sim$ 5,000 galaxies. Including new photometric measurements for some 2000 galaxies, this catalogue additionally consists of measurements from the SCI catalogue \cite{G97a}, SFI, SC2, SF2, along with data published in Mathewson et al (1992) and Mathewson \& Ford (1996). Of the total sample of 4861 galaxies for which reliable Tully-Fisher parameter measurements were obtained, 807 galaxies across 31 clusters were deemed suitable by \citeauthor{Spring07} for Tully-Fisher and H$_{0}$ measurement applications. Included in this catalogue are measurements of the recessional velocities, V$_{\mathrm{CMB}}$, for each galaxy, where these velocities have been transformed to the CMB reference frame to correct for motions relative to the CMB to approximate Hubble flow.\\

In order to maintain consistent treatment between the calibration and cluster data samples, we adopted from this catalogue the galaxies whose line widths has been measured by \cite{Court09}. In emphasising this consistency, we reject galaxies for which W$_{m50}$ line widths were not available in \citeauthor{Court09}. Line widths were assigned measurement errors as a function of their signal to noise, where an error of 20 km s$^{-1}$ or less signals a profile that is appropriate for Tully-Fisher applications. As a preliminary cut to the data, all line width profiles with an assigned error larger than 20 km s$^{-1}$ were rejected.\\

We adopt all other measurements from the SFI++ catalogue where used, apart from the line widths for the reasons outlined above. Photometry and line width measurements for these galaxies are corrected in a similar manner to that of the calibration sample galaxies, given by equations (2) and (3) respectively.

\subsection{Selection Criteria}
Following \cite{sa00}, we apply the following cuts to the cluster data:
\begin{enumerate}
 \item Galaxies with inclinations $i$ $\leq$ 40$^{\circ}$ are rejected as uncertainties in the inclination measurement begin to dominate line width errors. The deprojection of the line width to an edge-on measurement becomes increasingly large and uncertain as the inclination angle decreases. 
\item Both the calibration sample and the cluster should encompass the same rotational velocity distribution. Since the lowest rotational velocities of the calibration sample approach 180 km s$^{-1}$, we reject galaxies from the cluster sample with smaller velocities, as indicated in Figure 5.

\item An upper limit to the internal extinction correction of 0.75 mag is applied, corresponding to highly inclined galaxies which have more than half of their light is subject to self-scattering.
\end{enumerate}

Regarding the internal extinction cut-off, \citeauthor{sa00} emphasise that the distribution of internal extinction values of the cluster data sample should correspond to that of the calibration sample. In reproducing their Figure 8 for our own data samples in Figure 6, we can see that an extinction cut-off to 0.75 mag remains appropriate.\\

Lastly, only clusters with 5 or more galaxies and with V$_{\mathrm{CMB}} \geq$ 2000 km s$^{-1}$ were used in the determination of H$_{0}$. This final sample, referred to as the Hubble Sample, consists of 261 galaxies across 15 clusters.

\subsection{Distances and Recessional Velocities for each Cluster}

For each cluster, individual distance measurements to the constituent galaxies were obtained by applying the Tully-Fisher parameters to their corrected line widths to yield absolute magnitudes, from which distance moduli could be determined in conjunction with the SFI++ apparent magnitude measurements. The cluster distance modulus is then simply taken as the mean of these distances, with the RMS dispersion as the corresponding uncertainty.\\

From the recessional velocities of each individual galaxy, we can determine an overall recessional velocity for the cluster by taking the mean of these values. Whilst an uncertainty can be similarly adopted as the RMS dispersion about the mean of these values, we instead adopt a larger uncertainty of 300 km s$^{-1}$ for each cluster's recessional velocity. This uncertainty corresponds to the dispersion of peculiar velocities for clusters as found by \cite{Giov98}, and allows us to account for any large scale peculiar motions and anisotropy in the distribution of clusters across the sky.\\

Both of these quantities are of course dependent on the assumption that the cluster is a self-contained object with little substructure. This can be verified for each cluster by plotting histograms of distance moduli and recessional velocity measurements, where we expect these quantities to be normally distributed about some mean value. The presence of outliers are not likely to be indicators of measurement errors, but instead may point to errors in the method by which galaxies are assigned to a particular cluster. For the galaxies in the SFI++ template sample, cluster membership assignment follows that of \citet{G97a}, in which galaxies are assigned membership based on spatial proximity to a clearly defined central region, and consistency with recessional velocity measurements. Galaxies that satisfy both of these criteria for a particular cluster constitute the \textit{in} sample. Galaxies that do not satisfy spatial proximity, but have a redshift that matches that of the assigned members form a combined sample with the \textit{in} galaxies, referred to as the \textit{in+} sample \citep{G97a}. For the sake of preserving as large a sample as possible, we have utilised the \textit{in+} sample in this study.\\

Nevertheless, since we are relying on \textit{in+} sample, our cluster sample is more prone to outliers as we limit the sample to galaxies for which corrected magnitudes can be reliably determined. This becomes particularly problematic for galaxies with less than 10 galaxies, from which a mean distance for the cluster must be measured. However, for clusters with fewer galaxies, the random error measurement on the distance will be larger, and hence suitably reflect the loose constraint.\\

With respect to measuring a mean cluster recessional velocity, we are only limited to galaxies that have a reliable cluster assignment, since the redshift for each galaxy can be accurately determined, independent of the properties that limit our sample by suitability of magnitude corrections. Again however, this method relies on the assumption that the cluster is a gravitationally bound system supported by the random motion of its galaxies. Under this assumption, the line of sight motion of each galaxy varies about some mean recessional motion of the cluster corresponding to random motions undergone within the cluster. Any evidence of substructure, in which distinct groups that have yet to coalesce would have their own mean $cz$ values, would be revealed by a $cz$ histogram with multiple distinct peaks.\\

For this study, recessional velocity measurements for each cluster have been determined using the entire SFI++ template sample, which were then compared with values published in \cite{MA06} and \cite{G97b}. This is to ensure that the sample from which we are working is suitably large enough for accurate measurements to be made. The immediate limitation of course in using only these galaxies is that we have only sampled the spiral galaxies in each cluster, and ignored the large number of elliptical galaxies. We may be able to take some solace in the fact that since we are sampling spiral galaxies only, which are found in greater numbers in the outer regions of most clusters, we avoid the large infall motions of galaxies closer to cluster cores which can induce large dispersions in an overall $cz$ measurement \citep{mo00}. All our measurements are consistent with published values, apart from cluster A2634, where we instead adopt the published value in favour of our own. For A2634, our measurement is based on 22 galaxies, whereas the value from \citeauthor{MA06} is based on 200 galaxies, thereby providing a better sample for analysis.\\

Final distances, recessional velocities, and values for H$_{0}$ for each cluster are presented in Table 3. For each parameter, 1$\sigma$ random errors are included.


\section{Hubble Diagram and the Value of the Hubble Constant}

With distance and recessional velocity measurements, H$_{0}$ values can be determined for each individual cluster, and a weighted mean can be adopted, where the random errors are used as the weights. For the data sample presented in Figure 7, the value of H$_{0}$ arrived at via a TRGB-calibrated Tully-Fisher relation is 79 $\pm$ 2 (random) km s$^{-1}$ Mpc$^{-1}$.

We can further construct a Hubble Diagram (Figure 8).

\subsection{ Error Propagation and a Final Value for H$_{0}$}
Random and systematic errors are propagated independently into two final uncertainties for H$_{0}$. Random errors can be reduced statistically with relative ease, but systematic errors cannot be reduced in this manner, and hence require refinement of existing methods, or new methods altogether. These errors must be tracked carefully through each stage of the calculation from the calibration sample magnitudes and line widths until a value of H$_{0}$ is derived.\\

When determining a total error for a particular quantity in which we are propagating multiple individual errors, these are typically added in quadrature since these measurement uncertainties are independent of one another. The linear addition of errors instead indicates a correlation between errors, such that an overestimate of one quantity implies an overestimate in the other. But where error measurements are independent, quadrature addition of uncertainties provides a smaller, and more appropriate overall uncertainty \citep{Taylor97}.\\

The following error propagation recipe follows closely that of \cite{sa00}. 

\subsubsection{Errors in the Calibration Sale} 
Random errors in the calibration scale are propagated as an uncertainty in H$_{0}$ through the intrinsic dispersion of the Tully-Fisher relation. Systematic errors arise in the calibration scale through the zero-point error in M$_{I}^{TRGB}$, and the zero-point uncertainty in the Tully-Fisher relation. As determined by \cite{ri07}, the systematic uncertainty in M$_{I}^{TRGB}$ is 0.02 mag.

\subsubsection{Random Errors in the Tully-Fisher Parameters}
Random errors in the corrected apparent magnitude are given by the quadrature addition of the uncertainty in the measurement and the internal extinction correction, the latter of which propagates the errors in the ellipticity and $\gamma$ parameters. As per \cite{G97a}, we assume an uncertainty of 25\% for $\gamma$. Uncertainties in the redshift correction and galactic extinction are sufficiently small so as to be ignored. The final expression for the error in m$_c$ is given by:
\begin{equation}
\epsilon_{m}^{2} = \epsilon_{obs}^2 + (0.25 \mathrm{log}(1 - e))^{2} + \left(\frac{0.434 \gamma}{(1-e)} \epsilon_{e}\right)^{2} 
\end{equation}

Errors in the corrected line widths are purely random, and are given by the quadrature addition of the uncertainty in the measurement and the inclination:

\begin{equation}
\epsilon_{W}^{2} = \left(\frac{1}{(1+z)\mathrm{sin}i}\epsilon_{W,meas}\right)^{2} +\left(\frac{-W_{obs}\mathrm{cos}i}{(1+z)(\mathrm{sin}i)^{2}}\epsilon_i\right)^{2}
\end{equation}

where the uncertainty in the inclination, $\epsilon_{i}$ is the propagated uncertainty in the ellipticity:

\begin{equation}
\epsilon_{i}=\frac{1-e}{(1-q_{0}^{2})\left[1-\frac{(1-e)^{2}-q_{0}^2}{1-q_{0}^2}\right]^{\frac{1}{2}}\left[\frac{(1-e)^{2}-q_{0}^2}{1-q_{0}^{2}}\right]^{\frac{1}{2}}}\epsilon_{e}
\end{equation}

The uncertainty in the ellipticity is given by the empirically determined relation, $\epsilon_{e} = 0.09 -0.12e + 0.037e^{2}$, as presented in \cite{G97b}. It should be noted that this evaluation of $\epsilon_{e}$ has the potential to underestimate the error for low ellipticities, but for the most part these are not present in the sample, since low ellipticities translate to approximately face-on galaxies which are excluded from the sample. Thus, a final error on logW is given by: 
\begin{equation}
\epsilon_{logW}=\frac{\epsilon_{W}}{0.434 W_{c}}
\end{equation}

\subsubsection{Random Errors in the Tully-Fisher Distance Moduli and H$_{0}$}
For each individual galaxy, the random error in the distance modulus is the quadrature addition of the errors in the apparent magnitude measurement and the random error in the absolute magnitude. The random error in the absolute magnitude consists of the uncertainties propagated by the Tully-Fisher relation: the intrinsic dispersion, $\sigma_{\mathrm{int}}$, and the line width error, where the error in logW is multiplied by the slope of the Tully-Fisher relation, $b$, to be expressed in magnitudes.
\begin{equation}
\epsilon_{\mu,rand}^{2} = \epsilon_{m}^{2}+b\epsilon_{logW}^{2}+\sigma_{\mathrm{int}}^{2}
\end{equation}
For each cluster, a mean distance modulus is adopted as the cluster distance, for which the uncertainty is simply taken as the RMS dispersion. There is a 1/$\sqrt{n}$ dependence on this error, and hence this random error component is reducible for clusters with a larger number of galaxies.\\ 

For individual cluster measurements, the random error in H$_{0}$ is comprised of the errors in the distance scale and the error in the adopted recessional velocity, $\epsilon_{cz}$, of the cluster. The uncertainty on the measured $cz$ is simply the dispersion of the individual member galaxy $cz$ values, divided by the square root of the number of galaxies in the cluster. However, this uncertainty reflects only the extent to which the averaged motion of the galaxies represents motion of the cluster as a whole. As per \citet{MA06}, we adopt an uncertainty of 300 km s$^{-1}$ for the recessional velocity of each cluster to account for the peculiar motion of each cluster, which is in accordance with \citet{Giov98}. The error in H$_{0}$ for a particular cluster is then given by:
\begin{equation}
\epsilon_{H,rand}^{2}=\left(\frac{\epsilon_{cz}}{d}\right)^{2}+(0.46\mathrm{H}_{0}\epsilon_{\mu,rand})^{2}
\end{equation}
where $d$ is the distance to the cluster in Mpc, determined from the average distance modulus.\\

We adopt as a final random error for H$_{0}$ as the RMS dispersion about the weighted mean, reduced by a factor of 1/$\sqrt{n}$, where we have used $n = 15$ individual measurements of H$_{0}$ taken from 15 clusters. 

\subsubsection{Systematic Uncertainty in H$_{0}$}
The systematic error in the calibration scale, consisting of the zero-point error on M$_{I}^{TRGB}$ and the zero point error in the Tully-Fisher relation propagate in to a final systematic uncertainty in H$_{0}$ as a systematic distance error. This is defined as:
\begin{equation}
\epsilon_{H,syst}=0.46 \mathrm{H}_{0} \epsilon_{\mu,syst}
\end{equation}

\subsection{A Final Measurement and Uncertainty for H$_{0}$}
Following the above procedures, we determine our final value for H$_{0}$:
\begin{center}
H$_{0}$ = 79 $\pm$ 2 (random) $\pm$ 3 (syst.) km s$^{-1}$ Mpc$^{-1}$
\end{center}

\subsection{The Sensitivity of H$_{0}$ to Changes in the Tully-Fisher Parameters}
Our least squares regression fits to our calibration sample are of course susceptible to the small number of calibrating points, where for such small samples outliers have a significant impact on the fit. As such, were this calibration sample to be extended, we would expect the slope and zero-point to change non-negligibly. Therefore, it is important to assess how changes in these Tully-Fisher parameters affect our final value for H$_{0}$. Using the values of the slope and zero-point for the bivariate fit, we evaluate the change in H$_{0}$ for a 3$\sigma$ change in each\footnote{The 3$\sigma$ do not represent confidence intervals, but simply possible variations.}, in order to anticipate possible values of H$_{0}$ that could be achieved with a more robust fit. For a $\pm$ 3$\sigma$ change in the slope, our value of H$_{0}$ changes by 2\%, and for a $\pm$ 3$\sigma$ in the zero-point, H$_{0}$ changes by 9\%. As can be seen by this quick analysis, the final value of H$_{0}$ exhibits minimal sensitivity to the slope, but a much stronger sensitivity to the zero-point.


\section{Conclusion}
In this paper, we have utilised the TRGB standard candle to construct a distance scale by which the expansion rate of the universe, or the Hubble constant, is measured. By determining apparent magnitudes for 11 galaxies with known TRGB distance moduli using observations from the SDSS catalogue, the ANU SkyMapper telescope along with data adopted from the literature, we have constructed a Cepheid-independent I-band Tully-Fisher relation:
\begin{equation}
M_{I} = (-7.81 \pm 0.65)(\mathrm{log}W - 2.5) - (21.11 \pm 0.07).
\end{equation}

This Tully-Fisher relation was then applied to the SFI++ Tully-Fisher template data from \cite{Spring07}, which initially consisted of 807 galaxies across 31 clusters. After applying cuts for line width and line width measurement error, internal extinction, inclination, and a minimum cluster number, we maintained a sample of 261 galaxies across 15 clusters. Applying the Tully-Fisher relation to line widths of these galaxies yielded absolute magnitudes for each galaxy which, when combined with their listed apparent magnitudes, gave a distance to each galaxy. For each cluster, an average distance could then be obtained, in addition to an average recessional velocity from measurements of individual galaxy redshifts. These two measurements provide the basis for measuring the Hubble constant via Hubble's Law, $cz = \mathrm{H}_{0}r$.\\

Random and systematic errors have been propagated through each step of the distance scale into a final uncertainty for H$_{0}$. Systematic errors continue to dominate distance-scale based measurements of H$_{0}$, which are principally introduced in the calibration scale, although we have reduced the systematic error effectively using the well constrained absolute magnitude of the TRGB in favour of the Cepheid distance scale. The final value of H$_{0}$ is taken to be 79 $\pm$ 2 (rand.) $\pm$ 3 (syst.) km s$^{-1}$ Mpc$^{-1}$, which represents an uncertainty of 4\%. While our measurement of H$_{0}$ represents a upper estimate relative to other recent determinations (74.2 $\pm$ 3.6 km s$^{-1}$ Mpc$^{-1}$ \citep{riess09}; 73 $\pm$ 2 (rand.) $\pm$ 4 (syst.) km s$^{-1}$ Mpc$^{-1}$ \citep{frm10}; 73 $\pm$ 5 (rand.) km s$^{-1}$ Mpc$^{-1}$ \citep{ms08}), but it is still consistent given the uncertainties.\\

Agreement with the results of \cite{fr01} implies that the TRGB and Cepheid distance scales are consistent.

\acknowledgements
Thanks go to James Schombert for providing the ARCHANGEL Photometry Package, along with generous and timely support. 
A big thank you is also extended to Sean Crosby, Christina Magoulas, Ta\"issa Danilovich, and Loren Bruns Jr for their ongoing assistance. 
Special thanks to 
the whole SkyMapper team for making it all possible in the first place.
This research has made use of IRAF, which is distributed by NOAO. NOAO is
operated by AURA under a cooperative agreement with NSF.

\pagebreak 
\appendix
\leftline{\bf Appendix: Calibration of Photometric Data}
As outlined in \cite{Bess05}, the idea behind calibration is to simply place photometric measurements from multiple instruments onto a single scale.

\section{SkyMapper Images}

In calibrating the SkyMapper images, standards from the E1 region from \cite{Grah82} were observed only once on September 24, 2010. Three exposures each were taken of the field in SkyMapper's $g$ and $i$ bands, from which an average instrumental magnitude for each star was determined. Magnitudes were determined using the IRAF $phot$ package, 
and corrected for atmospheric extinction using the k-coefficients for the Siding Springs Observatory from \cite{HungBess00}.
These were science verification observations for SkyMapper hence the
limited calibration available.\\

Despite the small sample of standards, we have obtained a calibration applicable to a suitable range in color. Fitted to the data are least squares regressions, one minimising residuals in the offset, and the other minimising residuals in both the offset and the color. Errors in the color are the propagated errors from the individual magnitude measurements, where the errors in the instrumental magnitude are simply the RMS dispersion about the adopted mean since multiple measurements were available. Errors in the offset measurement are the quadrature addition of the errors in the I-band magnitudes from Graham and the errors on $i$. \\

Also superimposed on the plot is the weighted mean which, over the plotted range in color, is consistent with the least squares regressions to within 0.05 mag. Added to the fact that there is no clear trend with color, we adopt a constant offset allowing conversion between SkyMapper's instrumental $i$ magnitude and the standard I magnitude. Thus, we determine an overall calibration offset constant of - 3.31 $\pm$ 0.05 mag, where we have taken the RMS dispersion as the error. 


\section{SDSS Images}
Standard fields for the SDSS catalogue were obtained through the SDSS Data Archive Server\footnote{http://das.sdss.org} and I-band magnitudes from \cite{Land92}\footnote{http://www.cfht.hawaii.edu/ObsInfo/Standards/Landolt/} were used.  Atmospheric extinction coefficients for the Apache Point Observatory (where the SDSS observations were performed) are taken from \citet{Hogg01}.\\

In an identical manner to the SkyMapper calibration, Sloan instrumental colors $g- i$ for each star were plotted against an offset, $I - i$, in Figure 10. Superimposed on this plot are the weighted mean along with a least squares regression.\\

As this figure indicates, the mean is consistent with the least squares fits to 0.005 mag over the covered color range, such that we can suitably apply a mean offset of 2.39 $\pm$ 0.01 mag to calibrate all SDSS images.  

\pagebreak

\begin{deluxetable}{lcccccccccc}
\rotate
\tablecaption{\bf Calibration sample of galaxies}
\startdata
NGC& $\alpha$ & $\delta$ &$z$ & E(B-V) & T & e& W$_{m50}$&$\mu_{\mathrm{TRGB}}$&I\\
&(J2000)&(J2000)& & & & & &(km s$^{-1}$)&(mag)\\
(1)&(2)&(3)&(4)&(5)&(6)&(7)&(8)&(9)&(10)\\
55	&	00 14 53.60	&	-39 11 47.9	&	0.00043	&	0.013	&	9	&	0.87	&	181	$\pm$	13	&	26.62	$\pm$	0.03	&	7.09	$\pm$	0.05	\\
247	&	00 47 08.55	&	-20 45 37.4	&	0.00052	&	0.018	&	7	&	0.67	&	216	$\pm$	11	&	27.84	$\pm$	0.02	&	7.99	$\pm$	0.21	\\
253	&	00 47 33.12	&	-25 17 17.6	&	0.000811	&	0.02	&	5	&	0.78	&	428	$\pm$	18	&	27.83	$\pm$	0.02	&	5.53	$\pm$	0.07	\\
300	&	00 54 53.5	&	-37 41 04	&	0.00048	&	0.013	&	7	&	0.25 $^{a}$	&	155	$\pm$	19	&	26.59	$\pm$	0.06	&	7.3	$\pm$	0.05 $^{a}$	\\
891	&	 02 22 33.4	&	 +42 20 57	&	0.001761	&	0.065	&	3	&	0.77 $^{a}$	&	437	$\pm$	22	&	29.98	$\pm$	0.08	&	8.47	$\pm$	0.03 $^{a}$	\\
2403	&	07 36 51.4	&	+65 36 09	&	0.000437	&	0.04	&	6	&	0.47 $^{a}$	&	227	$\pm$	22	&	27.5	$\pm$	0.05	&	7.29	$\pm$	0.05 $^{a}$	\\
3351	&	10 43 57.70	&	+11 42 13.7	&	0.002595	&	0.028	&	3	&	0.28 	&	274	$\pm$	16	&	29.92	$\pm$	0.03	&	8.38	$\pm$	0.03	\\
3368	&	10 46 45.7	&	+11 49 12	&	0.002992	&	0.025	&	2	&	0.33 $^{a}$	&	340	$\pm$	9	&	29.3	$\pm$	0.06	&	8.06	$\pm$	0.05 $^{a}$	\\
3621	&	11 18 16.5	&	-32 48 51	&	0.002435	&	0.08	&	7	&	0.51 $^{a}$	&	282	$\pm$	13	&	29.08	$\pm$	0.06	&	8.43	$\pm$	0.05 $^{a}$	\\
3627	&	11 20 15.0	&	+12 59 30	&	0.002425	&	0.032	&	3	&	0.39 $^{a}$	&	345	$\pm$	15	&	29.59	$\pm$	0.09	&	7.71	$\pm$	0.05 $^{a}$	\\
4258	&	12 18 57.5	&	+47 18 14	&	0.001494	&	0.016	&	4	&	0.64 $^{a}$	&	385	$\pm$	20	&	29.41	$\pm$	0.04	&	7.22	$\pm$	0.05 $^{a}$	\\
4826	&	12 56 43.69	&	+21 40 57.5	&	0.001361	&	0.04	&	2	&	0.48 	&	319	$\pm$	16	&	28.2	$\pm$	0.03	&	7.21	$\pm$	0.03	\\
7793	&	23 57 49.83	&	-32 35 27.7	&	0.000757	&	0.018	&	7	&	0.34	&	177	$\pm$	13	&	27.79	$\pm$	0.08	&	7.91	$\pm$	0.07	\\
\enddata
\tablecomments{Presentation of the parameters for our calibration sample of galaxies. Values for ellipticity and measured I-band magnitude are from this paper unless where specified. Values marked with $^{a}$ are taken from \cite{TP00}.}
\tablecomments{In column (1), the galaxy's NGC number is listed. In columns (2) and (3) are the right ascension and declination of the object in J2000 coordinates. The redshift, foreground reddening value (taken from \citeauthor{Schlegel98}), and the morphological T code are listed in columns (4), (5), and (6). Ellipticity values are listed in column (7), with the corresponding source code. 
The measured rotational velocities adopted from \citeauthor{Court09} (apart from NGC 247, NGC 891 and NGC 2403) are listed in column (8). Column (9) contains the TRGB moduli as taken from the EDD (apart from NGC 3351). Column (10) contains the measured I-band magnitudes, with the corresponding source code.}
\end{deluxetable}



\begin{table}[h]
\leftline{\bf Table 2. Tully-Fisher parameters.}
\vskip 1 cm
\begin{center}
\begin{tabular}{lccccc}
\hline
\hline
Fit& $b$ & $a$ & $\sigma_{obs}$ \\
&&&(mag)\\
\hline
Direct & -7.37 $\pm$ 0.26 & -21.02 $\pm$ 0.03& 0.54\\ 
Inverse & -9.97 $\pm$ 0.67 & -21.04 $\pm$ 0.15 & 0.63\\ 
Bivariate & -7.91 $\pm$ 0.62 &  -21.11 $\pm$ 0.07 & 0.52\\
\hline
\hline
\end{tabular}
\end{center}
\end{table}

\begin{table}[h]
\leftline{\bf Table 3. Cluster data}
\vskip 1 cm
\begin{center}
\begin{tabular}{lccccc}
\hline
\hline
Cluster & N &$\mu$ & D & V$_{\mathrm{CMB}}$& V$_{\mathrm{CMB}}$/D\\
& &(mag)&(Mpc)&(km s$^{-1}$)&(km s$^{-1}$Mpc$^{-1}$)\\
\hline
N383	&	21	&	33.95	$\pm$	0.11	&	62	$\pm$	3	&	4718	&	77	$\pm$	6	\\
N507	&	13	&	33.90	$\pm$	0.13	&	60	$\pm$	4	&	4660	&	77	$\pm$	7	\\
A194	&	7	&	33.98	$\pm$	0.13	&	62	$\pm$	4	&	5002	&	80	$\pm$	7	\\
A262	&	32	&	34.01	$\pm$	0.09	&	63	$\pm$	3	&	4710	&	74	$\pm$	6	\\
A397	&	5	&	35.15	$\pm$	0.40	&	107	$\pm$	20	&	9746	&	91	$\pm$	17	\\
A400	&	41	&	34.63	$\pm$	0.05	&	85	$\pm$	2	&	7011	&	83	$\pm$	4	\\
Cancer	&	35	&	33.94	$\pm$	0.07	&	61	$\pm$	2	&	4976	&	81	$\pm$	6	\\
A779	&	12	&	34.98	$\pm$	0.10	&	99	$\pm$	4	&	7255	&	73	$\pm$	4	\\
Antlia	&	5	&	32.89	$\pm$	0.28	&	38	$\pm$	5	&	3142	&	83	$\pm$	13	\\
Hydra	&	9	&	33.59	$\pm$	0.10	&	52	$\pm$	2	&	4118	&	79	$\pm$	7	\\
A1367	&	16	&	34.65	$\pm$	0.10	&	85	$\pm$	4	&	6711	&	79	$\pm$	5	\\
Coma	&	27	&	34.40	$\pm$	0.08	&	76	$\pm$	3	&	7070	&	93	$\pm$	5	\\
A3574	&	5	&	33.83	$\pm$	0.29	&	58	$\pm$	8	&	4869	&	83	$\pm$	12	\\
Pegasus	&	20	&	33.49	$\pm$	0.10	&	50	$\pm$	2	&	3447	&	69	$\pm$	7	\\
A2634	&	13	&	35.18	$\pm$	0.09	&	108	$\pm$	5	&	8895	&	82	$\pm$	4	\\
\hline
\end{tabular}
\tablecomments{Presentation of the final Hubble sample cluster data. For each cluster, its average distance modulus and corresponding distance in Mpc, along with CMB recessional velocity and the resultant value of H$_{0}$ are listed.}
\end{center}
\end{table}
\pagebreak
\begin{figure}[h]
\begin{center}
\includegraphics[clip, width=\textwidth]{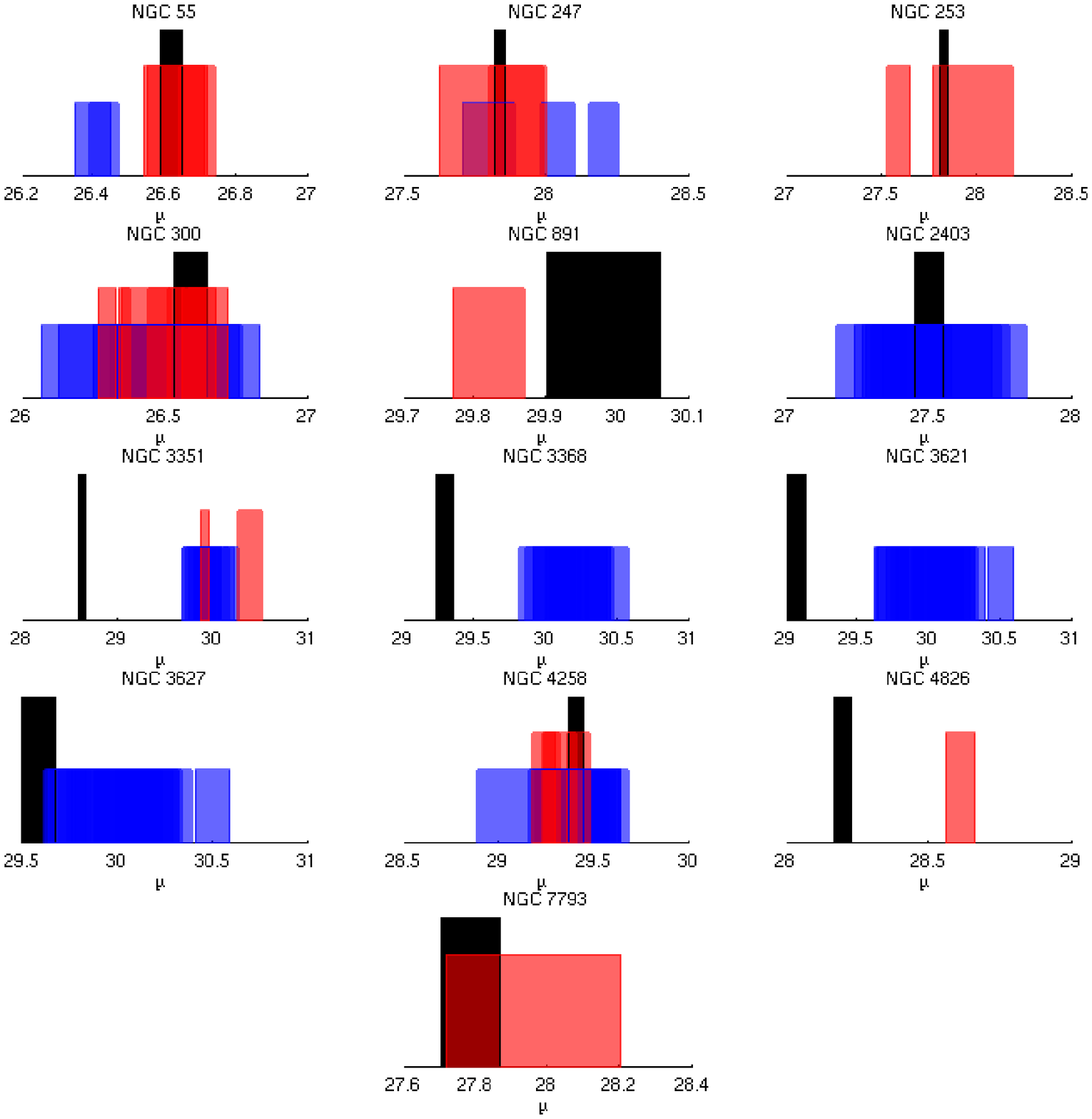}
\end{center}
\caption{Distribution of distance moduli measurements for each of our calibrator galaxies. Cepheid-based measurements are plotted in blue, TRGB-based measurements are plotted in red, and our adopted EDD TRGB values are plotted in black. Width corresponds to measurement error, and vertical scaling is arbitrary.}
\end{figure}
\begin{figure}[h]
\includegraphics[clip,width=\textwidth]{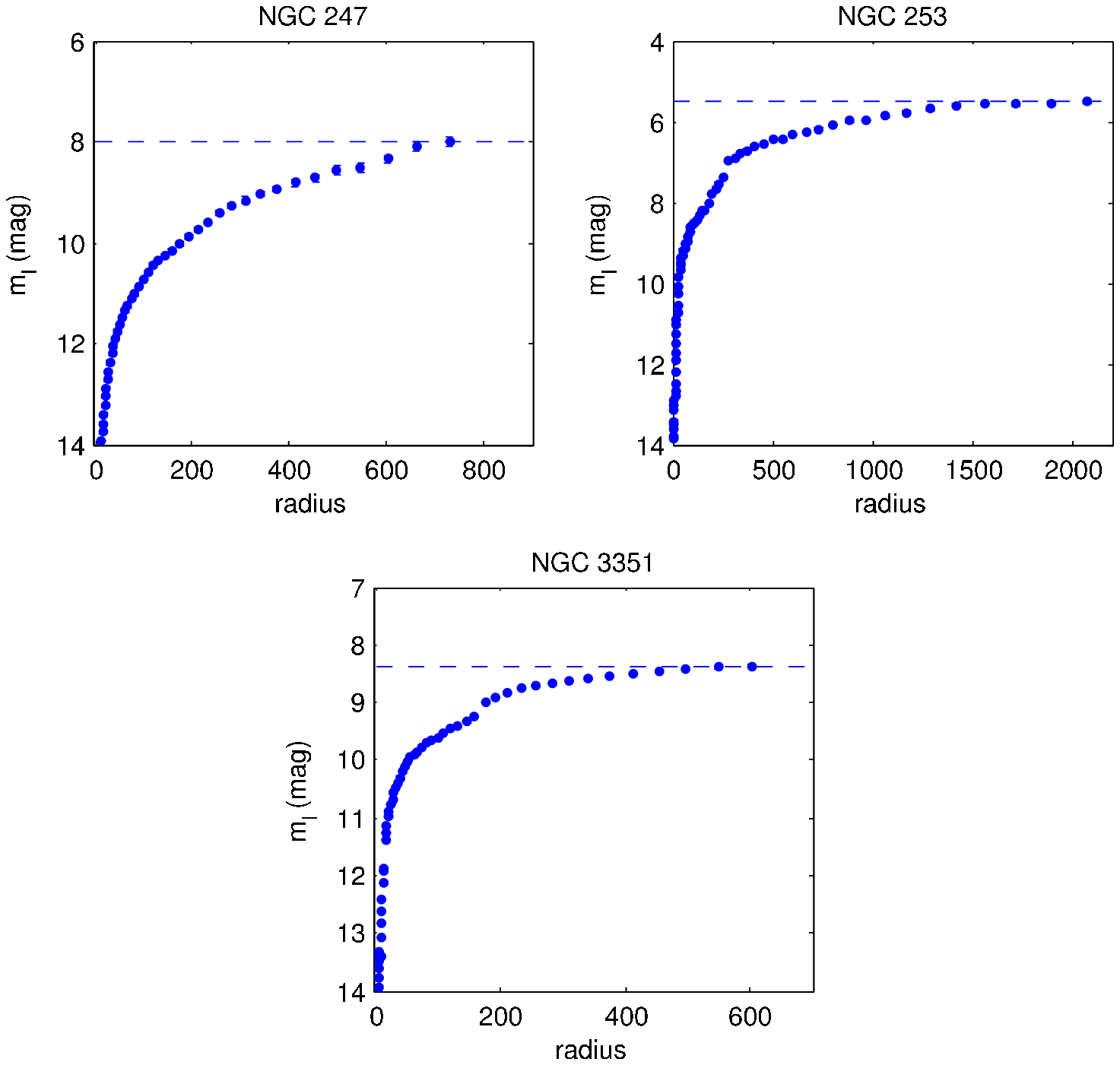}
\caption{Calibrated integrated magnitudes are plotted as a function of isophotal (semi-major) radius for each galaxy. The dashed line shows the final adopted magnitude. Fits for each galaxy were performed in IRAF.}
\end{figure}
\begin{figure}[h]
\includegraphics[clip,width=\textwidth]{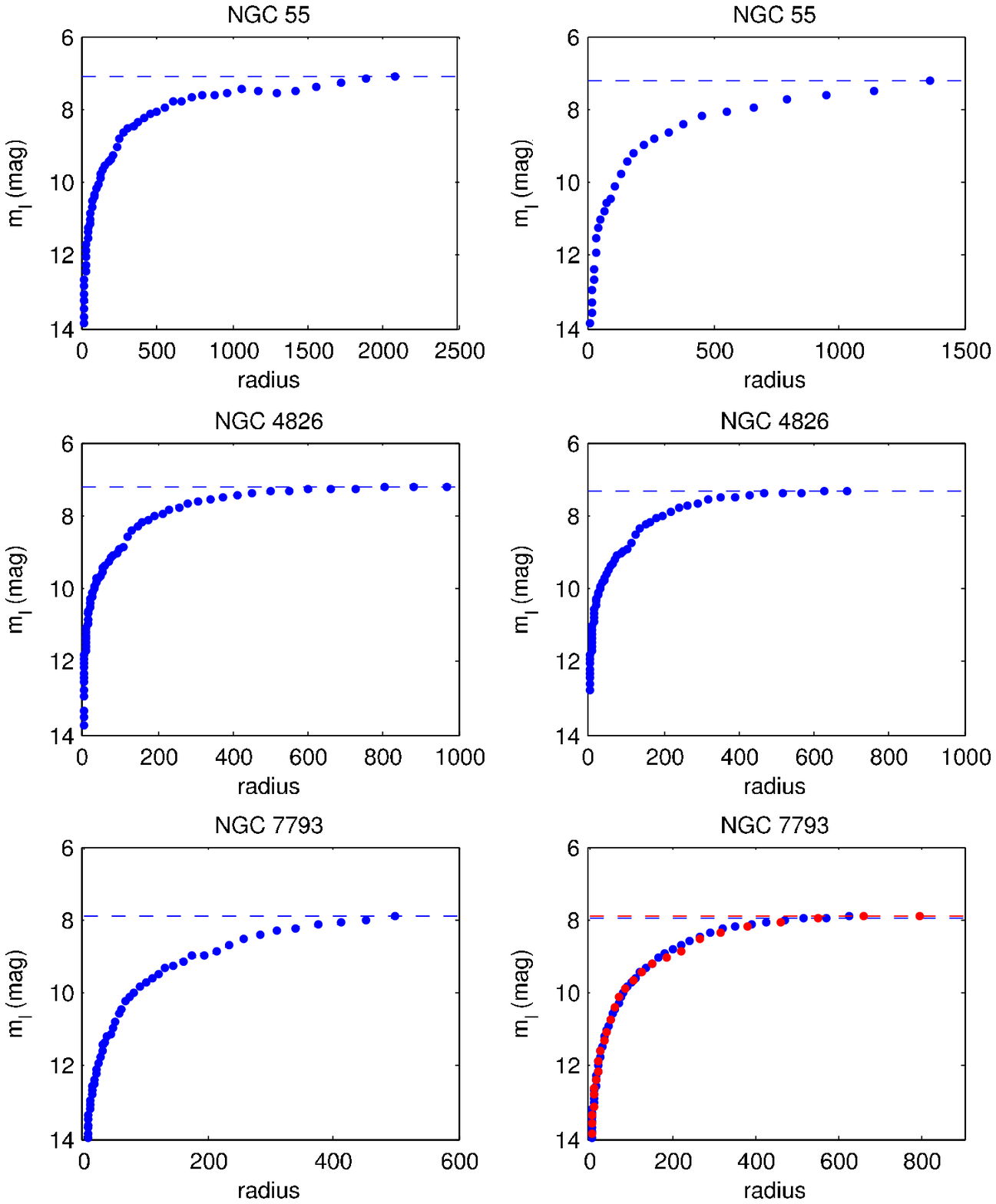}
\caption{Integrated magnitudes determined using IRAF are plotted on the left, with ARCHANGEL magnitudes on the right. The dashed line indicates the adopted raw magnitude. For NGC 7793, we have additionally plotted the integrated magnitudes determined using the ARCHANGEL $extremelsb$ fitting routine in red, which accommodates faint galaxies.}
\end{figure}
\begin{figure}[h]
\begin{center}
\includegraphics[clip, width=\textwidth]{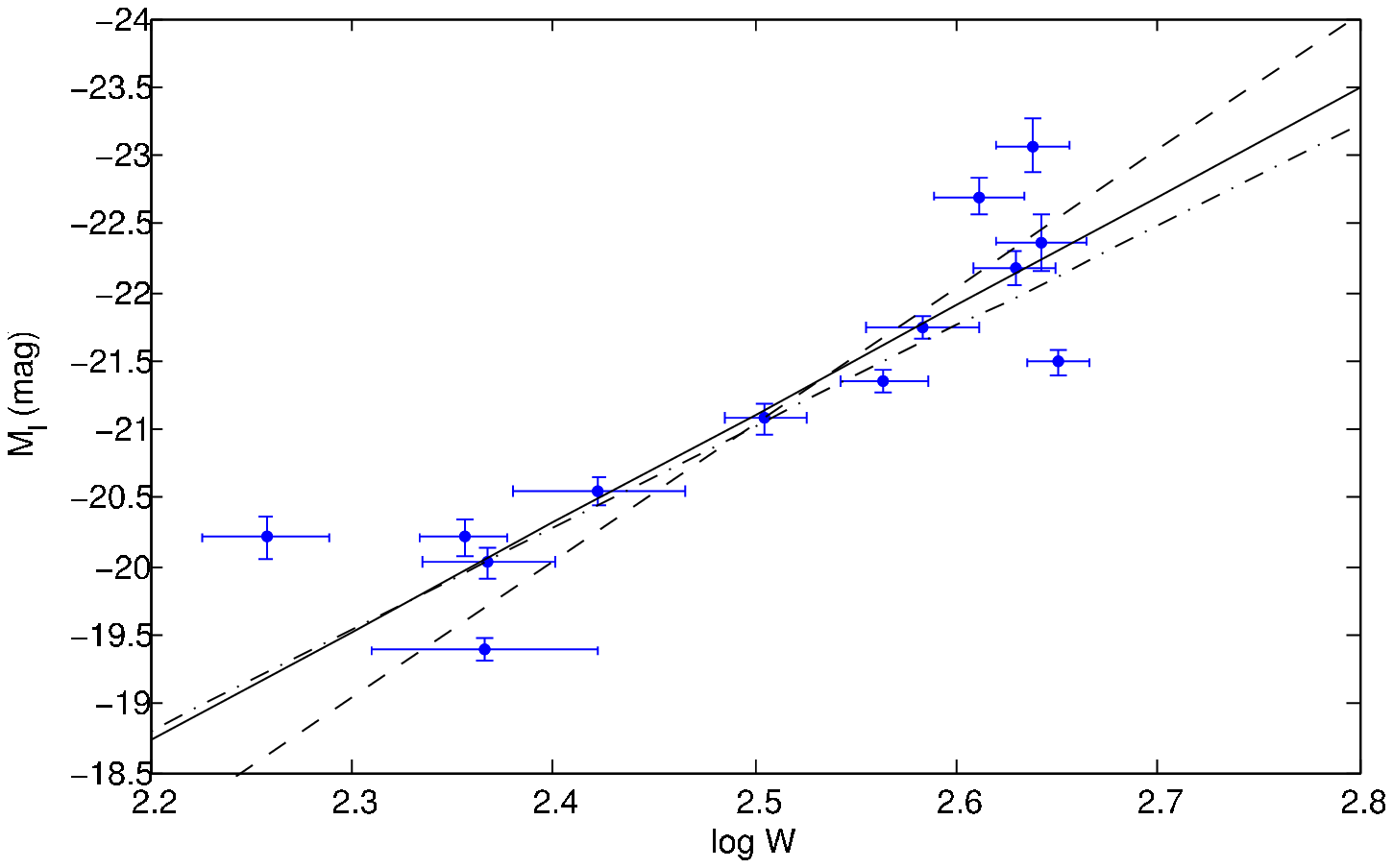}
\end{center}
\caption{Measured I-band Tully-Fisher Relation.}
\end{figure}
\begin{figure}[h]
\begin{center}
\includegraphics[clip, width=\textwidth]{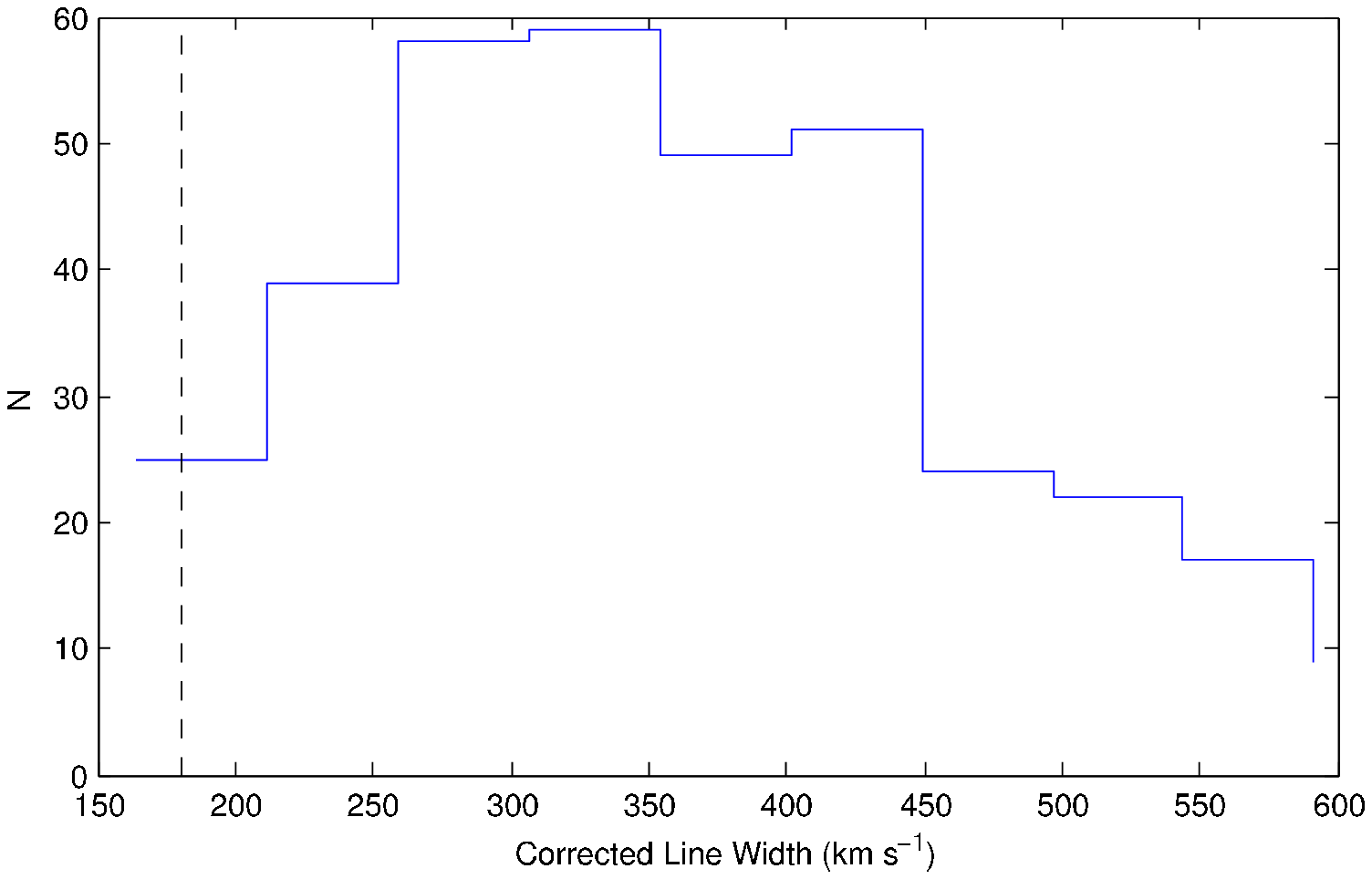}
\end{center}
\caption{Distribution of corrected line widths for the cluster sample. The vertical dashed line marks 180 km s$^{-1}$.}
\end{figure}
\begin{figure}[h]
\begin{center}
\includegraphics[clip, width=\textwidth]{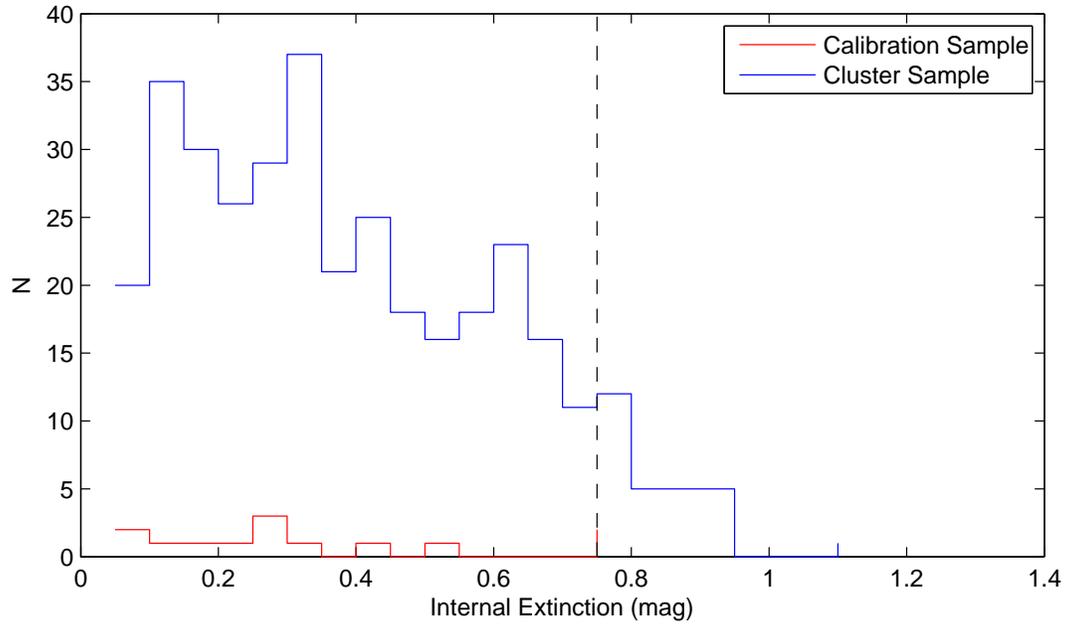}
\end{center}
\caption{Comparison of internal extinction distributions for the calibration and cluster samples.}
\end{figure}
\begin{figure}[h]
\begin{center}
\includegraphics[clip, width=\textwidth]{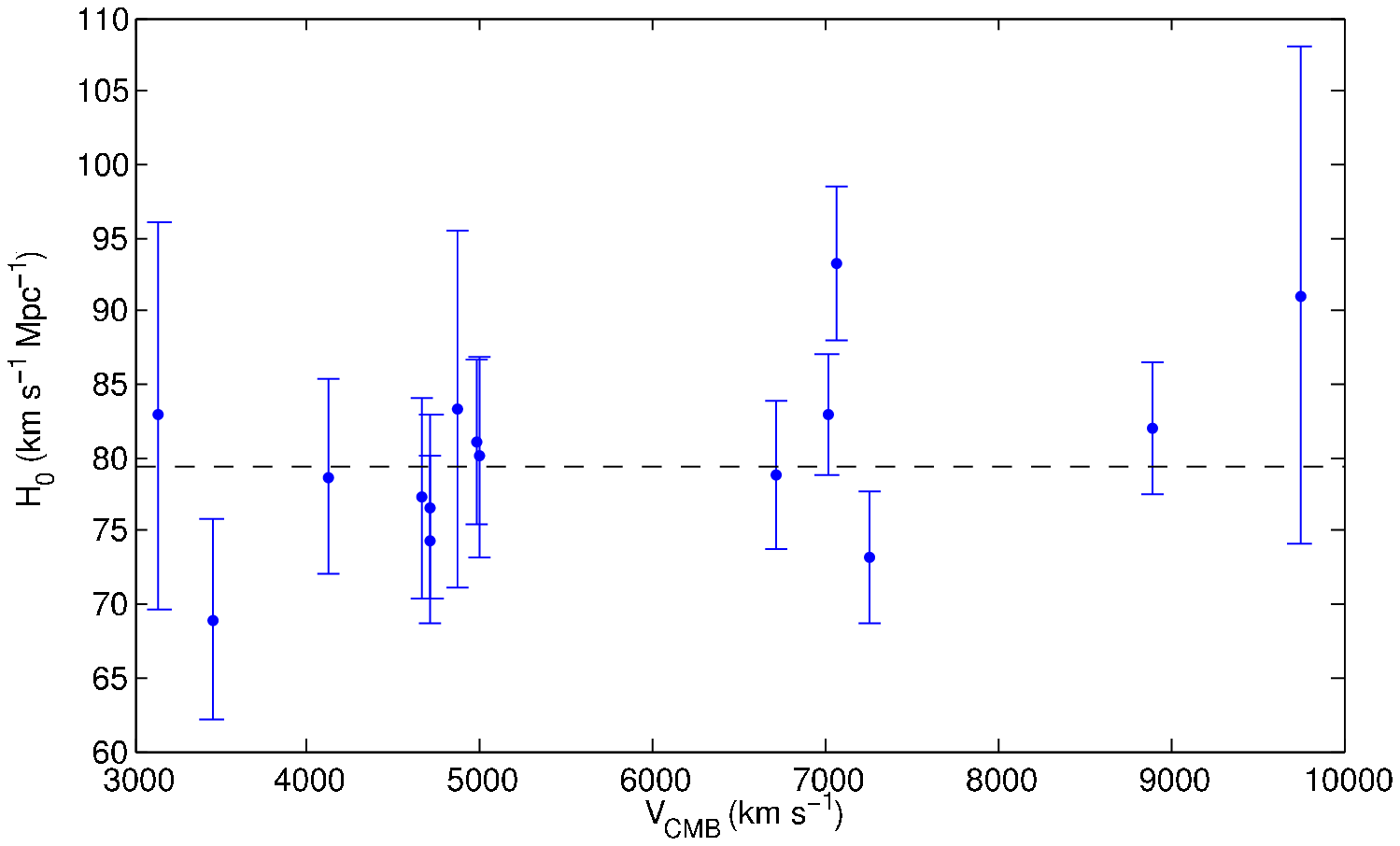}
\end{center}
\caption{Values for the Hubble constant for each individual cluster (with 1$\sigma$ random errors), where the weighted mean is plotted as the dashed line. These values are published in Table 3.}
\end{figure}
\begin{figure}[h]
\begin{center}
\includegraphics[clip, width=\textwidth]{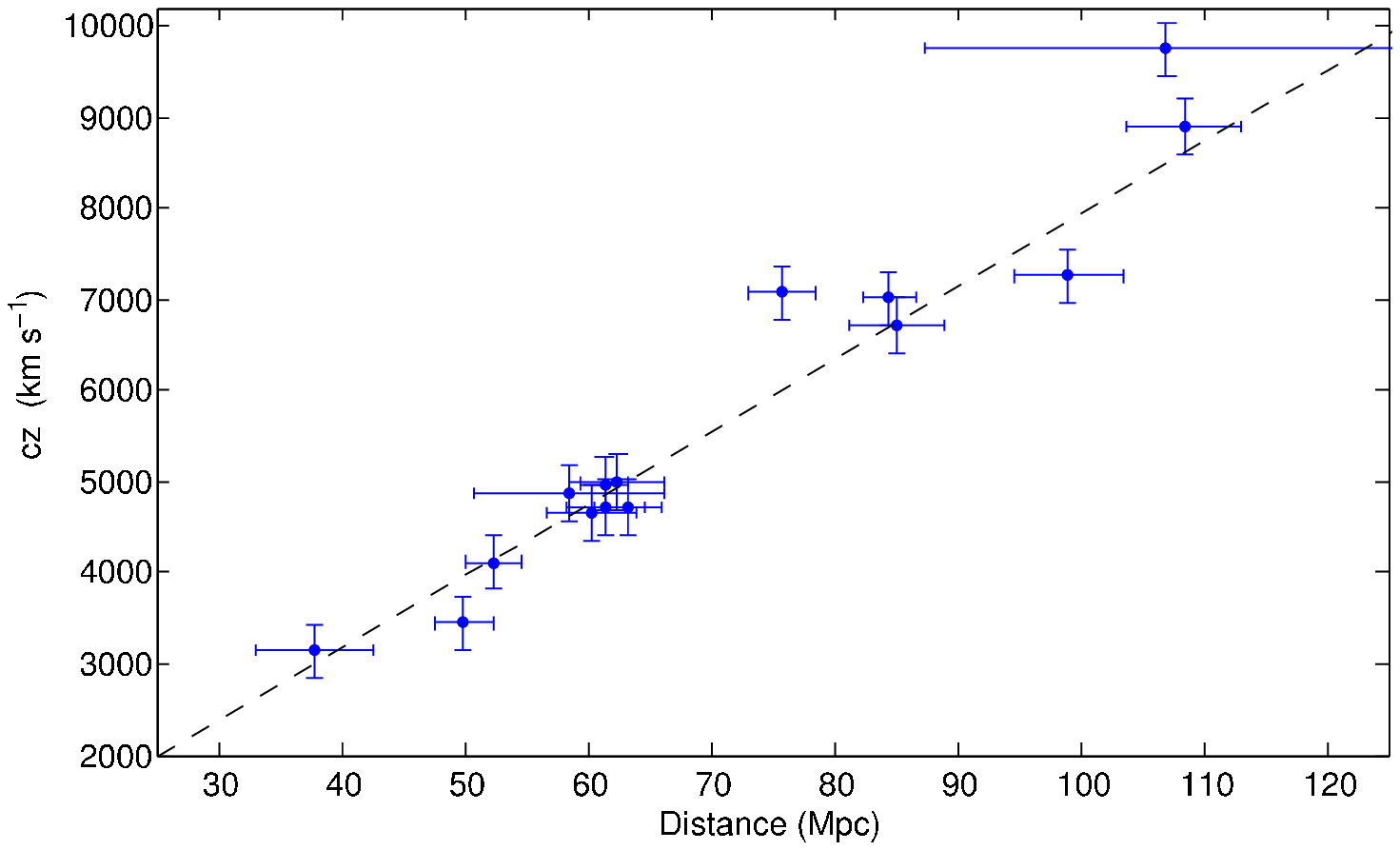}
\end{center}
\caption{Hubble diagram for the sample of clusters. The slope of the dashed line corresponds to the weighted mean for H$_{0}$.}
\end{figure}
\begin{figure}[h]
\begin{center}
\includegraphics[clip, width=\textwidth]{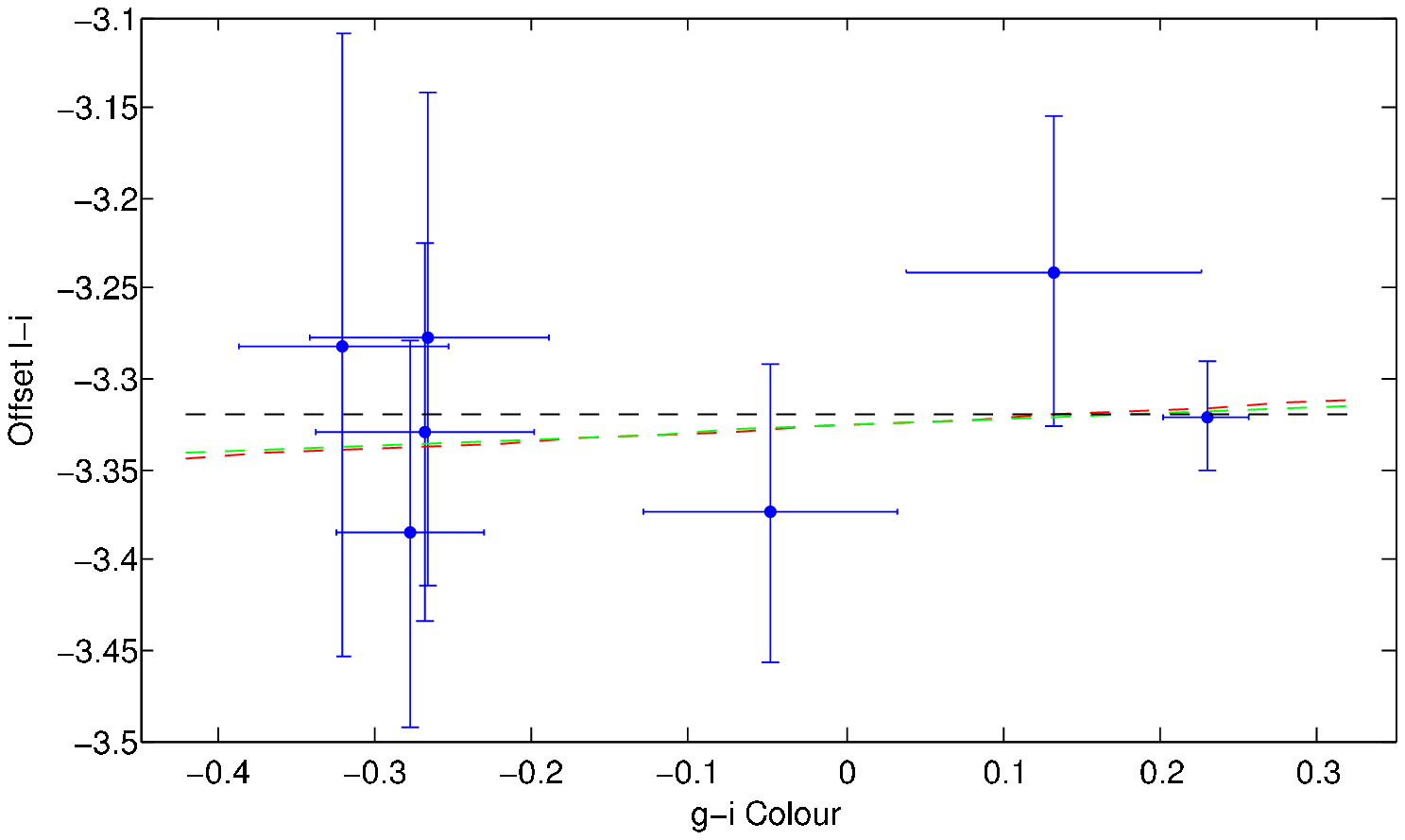} 
\end{center}
\caption{SkyMapper I-band calibration plot.}
\end{figure}
\begin{figure}[h]
\begin{center}
\includegraphics[clip, width=\textwidth]{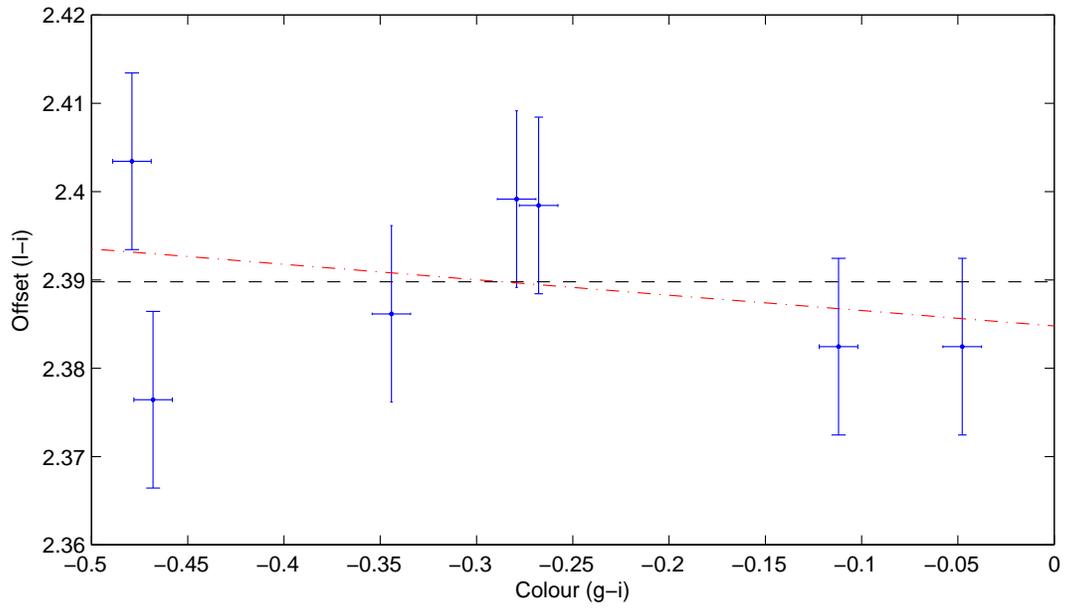}
\end{center}
\caption{SDSS I-band calibration plot.}
\end{figure}


\begin{thebibliography}{}
\bibitem[Bellazini, Ferraro, \& Pancini (2001)]{BEL01}Bellazini, M., Ferraro, F., \& Pancini, E. 2001,  \apj, 556, 535
\bibitem[Bessell (2005)]{Bess05}Bessell, M. 2005, \araa, 43, 293
\bibitem[Carignan \& Puche (1990)]{Carpu90}Carignan, C. \& Puche, D. 1990, \aj, 100, 641
\bibitem[Courtois \etal (2009)]{Court09}Courtois, H. \etal 2009, \aj, 138, 1938
\bibitem[Dalcanton \etal (2009)]{Dalcan09}Dalcanton, J. \etal 2009, \apjs, 183, 67
\bibitem[de Vaucouleurs \etal (1995)]{devauc95}de Vaucouleurs, G. \etal 1995, VizieR Online Data Catalog, 7155
\bibitem[Feigelson \& Babu (1992)]{FBab92}Feigelson, E. \& Babu, G. 1992, \apj, 397, 55
\bibitem[Ferrarese \etal (2000)]{fe00}Ferrarese, L. \etal 2000, \apj, 529, 745
\bibitem[Freedman \etal (2001)]{fr01}Freedman, W. \etal 2001, \apj, 553, 47
\bibitem[Freedman \& Madore (2010)]{frm10}Freedman, W. \& Madore, B. 2010, \araa, 48, 673
\bibitem[Gibson \etal (2000)]{gib00}Gibson, B. \etal 2000, \apj, 529, 723
\bibitem[Giovanelli \etal (1994)]{Giov94} Giovanelli, R. \etal 1994, \aj, 107, 2036
\bibitem[Giovanelli \etal (1997a)]{G97a} Giovanelli, R. \etal 1997a, \aj, 113, 22
\bibitem[Giovanelli \etal (1997b)]{G97b} Giovanelli, R. \etal 1997b, \aj, 113, 53
\bibitem[Giovanelli \etal (1998)]{Giov98} Giovanelli, R. \etal 2007, \aj, 116, 2632
\bibitem[Giovanelli \etal (2007)]{Giov07} Giovanelli, R. \etal 2007, \aj, 133 2569
\bibitem[Graham (1982)]{Grah82}Graham, J. 1982, \pasp, 94, 244
\bibitem[Han (1992)]{Han92}Han, M. 1992, \apjs, 81, 35
\bibitem[Hogg \etal (2001)]{Hogg01}Hogg, D., Finkbeiner, D., Schlegel, D., \& Gunn, J. 2001, \aj, 122, 2129
\bibitem[Huchtmeier \etal (2005)]{Hucht05}Huchtmeier, W. \etal 2005, \aap, 435, 459
\bibitem[Isobe \etal (1990)]{Isobe90}Isobe, T., Feigelson, E., Akritas, \& Babu, G. 1990, \aj, 364, 104
\bibitem[Keller \etal (2007)]{K07}Keller, S., Schmidt, B., Bessell,M., Conroy, P., Francis, P., Granlund, A., Kowald, E., Oates, A., 
Martin-Jones, T., Preston, T., Tisserand, P., Vaccarella, A. and Waterson,M. 2007, PASA, 24, 1
\bibitem[Jacobs \etal (2009)]{Jacobs09}Jacobs, B. \etal 2009, \aj, 138, 332
\bibitem[Kelson \etal (2000)]{ke00}Kelson, D. \etal 2000, \apj, 529, 768
\bibitem[Kennicutt, Freedman \& Mould (1995)]{ke95}Kennicutt, R., Freedman, W., \& Mould, J. 1995, \aj, 110, 1476
\bibitem[Kennicutt \etal (1998)]{ken98}Kennicutt, R. \etal 1998, \apj, 498, 181
\bibitem[Kent \etal (2008)]{Kent08}Kent, B. \etal 2008, \aj, 136, 713
\bibitem[Koribalski \etal (2004)]{Korib04}Koribalski, B. \etal 2004,\aj, 128, 16
\bibitem[Landolt (1992)]{Land92}Landolt, A. 1992, \aj, 116, 2632
\bibitem[Larson \etal (2010)]{la10}Larson \etal 2010, astro-ph 1001.4635
\bibitem[Lee, Freedman, \& Madore (1993)]{LFM93}Lee, M., Freedman, W. \& Madore, B. 1993, \apj, 417, 553
\bibitem[Macri \etal (2006)]{mac06}Macri, L. \etal 2006, \apj, 652, 1113
\bibitem[Madore \& Freedman (1995)]{MF95}Madore, B. \& Freedman , W. 1995, \aj, 109, 1645
\bibitem[Madore, Mager, \& Freedman (2009)]{Madmag09}Madore, B., Mager, V. \& Freedman, W. 2009, \apj, 690, 389
\bibitem[Makarov \etal (2006)]{MAK06}Makarov, D. \etal 2006, \aj, 132, 2729
\bibitem[Masters \etal (2006)]{MA06}Master, K., Springob, C., Haynes, M. \& Giovanelli, R. 2006, \apj, 653, 861
\bibitem[Mathewson, Ford \& Buchhorn (1992)]{dsm}Mathewson, D., Ford, V., \& Buchhorn, M. 1992, \apjs, 81, 413
\bibitem[Mathewson \& Ford (1996)]{mf}Mathewson, D. \& Ford, V. 1996, \apjs, 107, 97
\bibitem[Mendez \etal (2002)]{MEZ02}Mendez, B. \etal 2002, \aj, 124, 213
\bibitem[Meyer \etal (2008)]{MEY08}Meyer, M., Zwaan, M., Webster, R., Schneider, S. \& Staveley-Smith, L. 2008, \mnras, 391, 1712 
\bibitem[Mould \etal (2000)]{mo00}Mould, J. \etal 2000, \apj, 529, 786
\bibitem[Mould \& Sakai (2008)]{ms08}Mould, J. \& Sakai, S. 2008, \apjl, 686, L75, Paper I
\bibitem[Mould \& Sakai (2009a)]{ms09}Mould, J. \& Sakai, S. 2009a, \apj, 694,1331, Paper II
\bibitem[Mould \& Sakai (2009b)]{ms09b}Mould, J. \& Sakai, S. 2009b, \apj, 697, 996, Paper III
\bibitem[Pierce \& Tully (1992)]{tp92} Pierce, M. \& Tully, R. 1992, \apj, 387, 47
\bibitem[Press \etal (1992)]{Press92}Press, W., Teukolsky, S., Vetterling, \& Flannery, B. 1992, Numerical Recipes in C. Cambridge: CUP.
\bibitem[Riess \etal (2009)]{riess09}Riess, A. \etal 2009, \apj, 699, 539
\bibitem[Rizzi \etal (2007)]{ri07}Rizzi, L. \etal 2007, \apj, 661, 815
\bibitem[Rots (1980)]{Rots1980}Rots, A., 1980, \aaps, 41, 189
\bibitem[Saintonge \etal (2008)]{Saint08}Saintonge, A. \etal 2008, \aj, 135, 588
\bibitem[Sakai, Madore, \& Freedman (1996)]{SMF96}Sakai, S., Madore, B., \& Freedman, W. 1996, \apj, 461, 713
\bibitem[Sakai \etal (2004)]{sa04}Sakai, S., Ferrarese, L., Kennicutt, R., \& Saha, A. 2004, \apj, 608, 42 
\bibitem[Sakai \etal (2000)]{sa00}Sakai, S., \etal 2000, \apj, 529, 698
\bibitem[Salaris \& Girardi (2005)]{sg05}Salaris, M. \& Girardi, L. 2005, \mnras, 357, 669
\bibitem[Sandage \& Tammann (2006)]{str07}Sandage, A. \& Tammann, G. 2006, \araa, 44, 93
\bibitem[Sandage \& Tammann (2008)]{st08}Sandage, A. \& Tammann, G. 2008, \apj, 686, 779
\bibitem[Saviane \etal (2008)]{sav08} Saviane, I. \etal 2008, \apj, 678 179
\bibitem[Schombert (2007)]{Schomb07}Schombert, J. 2007, astro-ph 0703646 
\bibitem[Schlegel \etal (1998)]{Schlegel98}Schlegel, D. \etal 1998, \apj, 500, 525
\bibitem[Schweizer \etal (2008)]{schw08} Schweizer, F. \etal 2008, \aj, 136 1482
\bibitem[Springob \etal (2005)]{Spring05}Springob, C., Haynes, M., Giovanelli, R., \& Kent, S. 2005, \apjs, 160, 149
\bibitem[Springob \etal (2007)]{Spring07}Springob, C., Masters, K., Haynes, M., Giovanelli, R., \& Marinoni, C.. 2007, \apjs, 172, 599 
\bibitem[Sung \& Bessell (2000)]{HungBess00}Sung, H. \& Bessell, M. 2000, \pasa, 17, 244
\bibitem[Tammann \etal (2008)]{str08}Tammann, G., Sandage, A., \& Reindl, B. 2008, \apj, 679, 52
\bibitem[Taylor (1997)]{Taylor97}Taylor, J. 1997, Introduction to Error Analysis, The Study of Uncertainties in Physical Measurements, NY: University Science Books.
\bibitem[Theureau \etal (2006)]{Theureau06}Theureau, G., Martin, J., Cognard, I., \&  Borsenberger, J. 2006, ASP Conf. Ser. 351, 429
\bibitem[Tully \& Fisher (1977)]{TF77}Tully, R.B., \& Fisher, J. R. 1977, \aap, 54, 661
\bibitem[Tully \& Pierce (2000)]{TP00}Tully, R.B., \& Pierce, M. 2000, \apj, 533, 744
\bibitem[Tully \etal (1998)]{Tully98}Tully, R.B. \etal 1998, \aj, 115, 2264

\end{thebibliography}
\end{document}